\documentclass[english,sort&compress]{achemso}
\usepackage[T1]{fontenc}
\usepackage[latin9]{inputenc}
\usepackage{textcomp}
\usepackage{graphicx}
\usepackage[numbers]{natbib}
\usepackage{subscript}

\makeatletter

\title{Unraveling the 3D atomic structure of a suspended graphene/hBN van
der Waals heterostructure}
\author{Giacomo Argentero\textsuperscript{1}, Andreas Mittelberger\textsuperscript{1},
Mohammad Reza Ahmadpour Monazam\textsuperscript{1}, Yang Cao\textsuperscript{2},
Timothy J. Pennycook\textsuperscript{1}, Clemens Mangler\textsuperscript{1},
Christian Kramberger\textsuperscript{1}, Jani Kotakoski\textsuperscript{1},
A. K. Geim\textsuperscript{2,3}, Jannik C. Meyer\textsuperscript{1}}
\affiliation{\textsuperscript{1}University of Vienna, Faculty of Physics, Boltzmanngasse
5, 1090 Vienna, Austria\\
\textsuperscript{2}Centre for Mesoscience and Nanotechnology and
\textsuperscript{3}School of Physics and Astronomy, University of
Manchester, Manchester M13 9PL, UK}
\email{jannik.meyer@univie.ac.at}

\makeatother

\usepackage{babel}
\begin{document}
\begin{abstract}
In this work we demonstrate that a free-standing van der Waals heterostructure,
usually regarded as a flat object, can exhibit an intrinsic buckled
atomic structure resulting from the interaction between two layers
with a small lattice mismatch. We studied a freely suspended membrane
of well aligned graphene on a hexagonal boron nitride (hBN) monolayer
by transmission electron microscopy (TEM) and scanning TEM (STEM).
We developed a detection method in the STEM that is capable of recording
the direction of the scattered electron beam and that is extremely
sensitive to the local stacking of atoms. Comparison between experimental
data and simulated models shows that the heterostructure effectively
bends in the out-of-plane direction, producing an undulated structure
having a periodicity that matches the moiré wavelength. We attribute
this rippling to the interlayer interaction and also show how this
affects the intralayer strain in each layer. 
\end{abstract}
KEYWORDS: van der Waals heterostructures, graphene, hexagonal boron nitride, scanning transmission electron microscopy.

\vspace{1cm}
\noindent
Boosted by the growing family of two-dimensional (2D) crystals, the
study of van der Waals heterostructures\citep{Geim2013} has emerged
in the last couple of years as one of the most active fields of research
in the science of 2D materials. The interest in these materials can
be explained by the practically infinite combinations of elementary
monolayers that can be artificially stacked to create structures with
desired properties. Among heterostructures, graphene on hexagonal
boron nitride (hBN) is one of the most studied. Both crystals are
chemically inert, have the same crystal structure and their lattice
constants only differ by 1.8\%, making them an ideal match. Compared
to SiO\textsubscript{2}, hBN provides a flatter, cleaner and electronically
more homogeneous insulating substrate \citep{Dean2010,Decker2011,Xue2011}
and is now routinely used to manufacture high-performance devices.\citep{Dean2010,Gannett2016,Kim2016}
Besides serving as an excellent substrate, hBN also provides graphene
with a periodical potential that, in the case of carefully aligned
crystals, can lead to small commensurate areas.\citep{Woods2014}

In contrast to previous works, where thick (bulk) hBN was used as
a substrate for single-layer graphene, we investigate here a freely
suspended heterostructure consisting of monolayer graphene on monolayer
hBN. Since the sample does not have a rigid support, fundamental phenomena
governing the interaction between the two crystals can be accessible
in the absence of external perturbations. The structure is investigated
\textit{via} transmission electron microscopy (TEM) and scanning transmission
electron microscopy (STEM), using low electron energies (80 keV for
TEM and 60 keV for STEM) in order to minimize radiation damage.\citep{Susi16NC}
In our STEM investigation, we use a detection scheme that is very
sensitive to small local tilts of the sample, which allows us to obtain
the 3D shape of the heterostructure through a comparison to model
structures.

\begin{figure}
\includegraphics[width=0.7\textwidth]{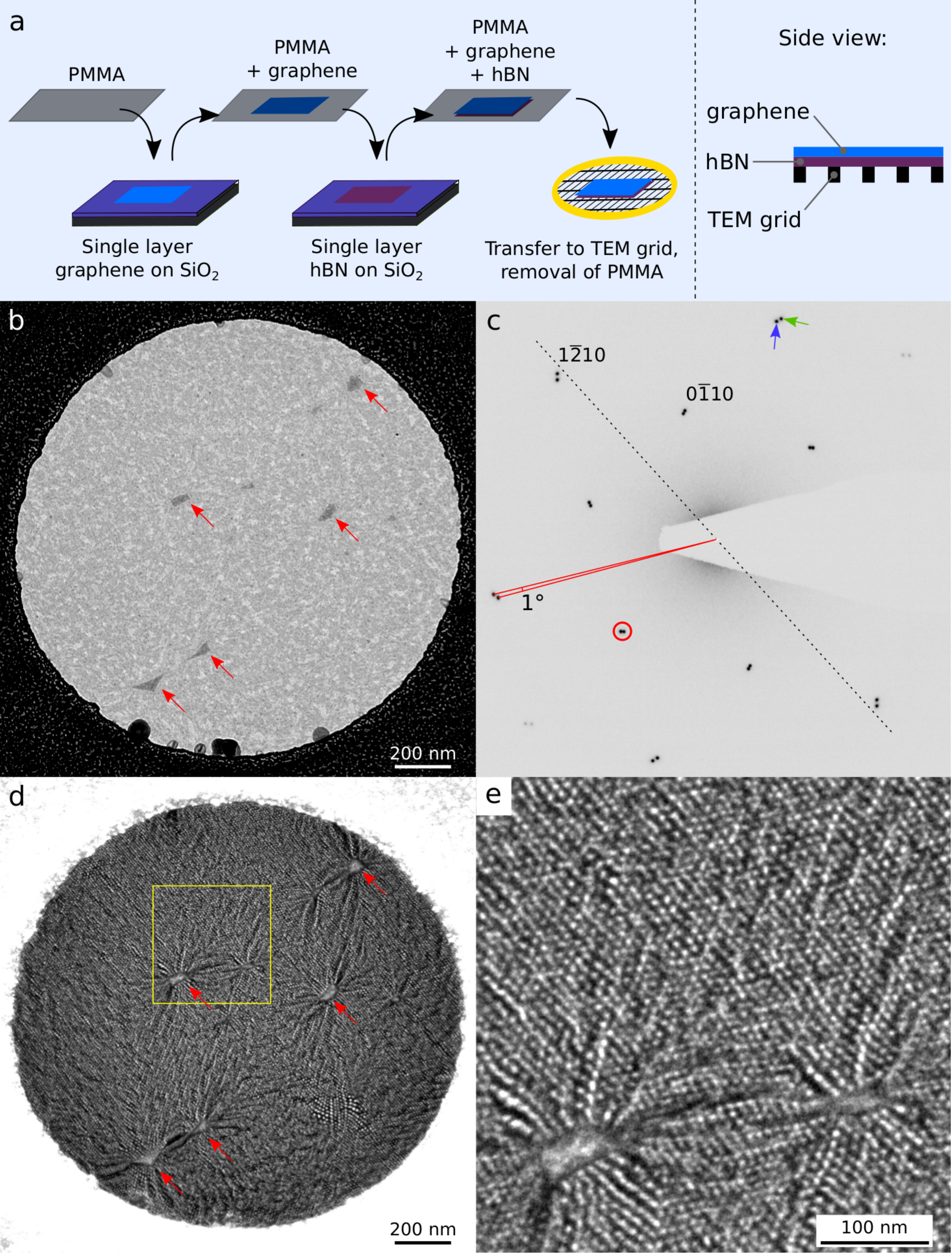}

\caption{CTEM analysis. The sample preparation and its transfer to the TEM
grid is schematically described in (a). (b) Bright field image of
the heterostructure freely suspended on a hole in the Quantifoil(R)
TEM grid. The red arrows indicate some aggregations of contaminants.
(c) Electron diffraction pattern of the heterostructure from the suspended
area in (b). The green and the blue arrows indicate one graphene and
one hBN diffraction spot, respectively. The misalignment between the
two crystals is 1\textdegree . With reference to dark field imaging
(d), the red circle marks the position of the objective aperture and
the dotted line indicates the tilt axis. (d) Dark field image of the
same area as in (b), acquired with a sample tilt of 16\textdegree .
The red arrows point to the same features as in (b). At these locations
the moiré interference pattern is completely suppressed. (e) Magnified
dark field image of the area inside the yellow square in (d).}
\label{ctem}

\end{figure}

To fabricate the sample we started with a freshly cleaved hBN crystal
on top of an oxidized Si wafer. A single layer hBN flake, chosen by
direct optical observation, was picked up by a single layer graphene
attached to a PMMA membrane following the method described in Ref.\ \citep{Kretinin2014}
and illustrated graphically in Fig.\ \ref{ctem}a. The dry transfer
method ensures an extremely clean interface between the flakes. Because
both hBN and graphene cleave preferentially along their main crystallographic
directions, during the transfer procedure we used flakes with well-defined
facets and aligned them (within a precision of 1.5\textdegree ) using
a rotating positioning stage under an optical microscope. The resulting
bilayer was then transferred onto a gold Quantifoil(R) TEM grid, where
portions of the heterostructure are freely suspended on holes measuring
approximately 1.5 \textmu m in diameter (see side view in Fig.\ \ref{ctem}a).
A thin layer of Pt was deposited on the TEM grids (prior to attaching
the 2D sample) in order to reduce hydrocarbon contamination.\citep{Longchamp2013}

The heterostructure was first investigated in a conventional TEM (Philips
CM200) operated at room temperature at 80 kV. Fig.\ \ref{ctem}b
shows a bright field (BF) image of the freely suspended heterostructure.
At the locations marked with red arrows, dark patches of dirt are
clearly visible. Their nature and position on the sample are discussed
below. Fig.\ \ref{ctem}c shows the electron diffraction pattern
obtained by illuminating the whole suspended area of the sample. Two
distinct sets of diffraction spots with hexagonal symmetry and 1\textdegree{}
relative angular rotation can be observed. Because of the mismatch
between the lattice constants of graphene and hBN (1.8\%, with hBN
being the larger of the two), it is possible to assign the outer and
the inner set of spots to the graphene and to the hBN lattice, respectively.
The green and the blue arrow in Fig.\ \ref{ctem}c indicate one graphene
and one hBN diffraction spot, respectively. The combination of lattice
mismatch and relative rotation is expected to produce a moiré superlattice
as already observed by AFM and STM in other works for the case of
graphene on bulk hBN.\citep{Yankowitz2012,Xue2011,Decker2011,Woods2014,Tang2013,Wang2016}
The moiré interference pattern can be conveniently visualized in the
TEM: Fig.\ \ref{ctem}d shows a dark field (DF) image of the same
area of Fig.\ \ref{ctem}b, acquired with a sample tilt of 16\textdegree .
Fig.\ \ref{ctem}e shows the area inside the yellow square in Fig.\ \ref{ctem}d
at higher magnification. The tilt axis in the reciprocal space is
indicated by the dotted black line in Fig.\ \ref{ctem}c and the
objective aperture used for DF imaging is marked by a red circle.
Note that the objective aperture was large enough to contain both
one graphene and one hBN spot. In contrast to the BF image, here a
strong modulation of the intensity appears, with bright spots arranged
in a triangular lattice. The periodicity of this modulation is 9.8
nm, very close to the predicted moiré period of 9.9 nm for a 1\textdegree{}
misaligned graphene/hBN bilayer.\citep{Hermann2012} Interestingly,
at some locations the moiré interference pattern is completely suppressed
(marked by arrows, corresponding to the same locations marked in Fig.\ \ref{ctem}b).
We interpret these regions (which also appear much darker in the BF
image) as pockets of contamination trapped between the layers, as
reported previously on the basis of cross-sectional TEM imaging.\citep{Haigh2012}
At these locations, the two layers are effectively separated by amorphous
contamination and the diffraction conditions for moiré interference
are suppressed. Indeed, the presence of the moiré in the DF-TEM images
indicates that most of the graphene/hBN interface is atomically clean.
Therefore, the clearly visible homogeneously distributed contamination,
similar to what is typically seen in TEM studies of graphene, must
be on the outer surfaces of the heterostructure.

\begin{figure}
\includegraphics[width=0.9\textwidth]{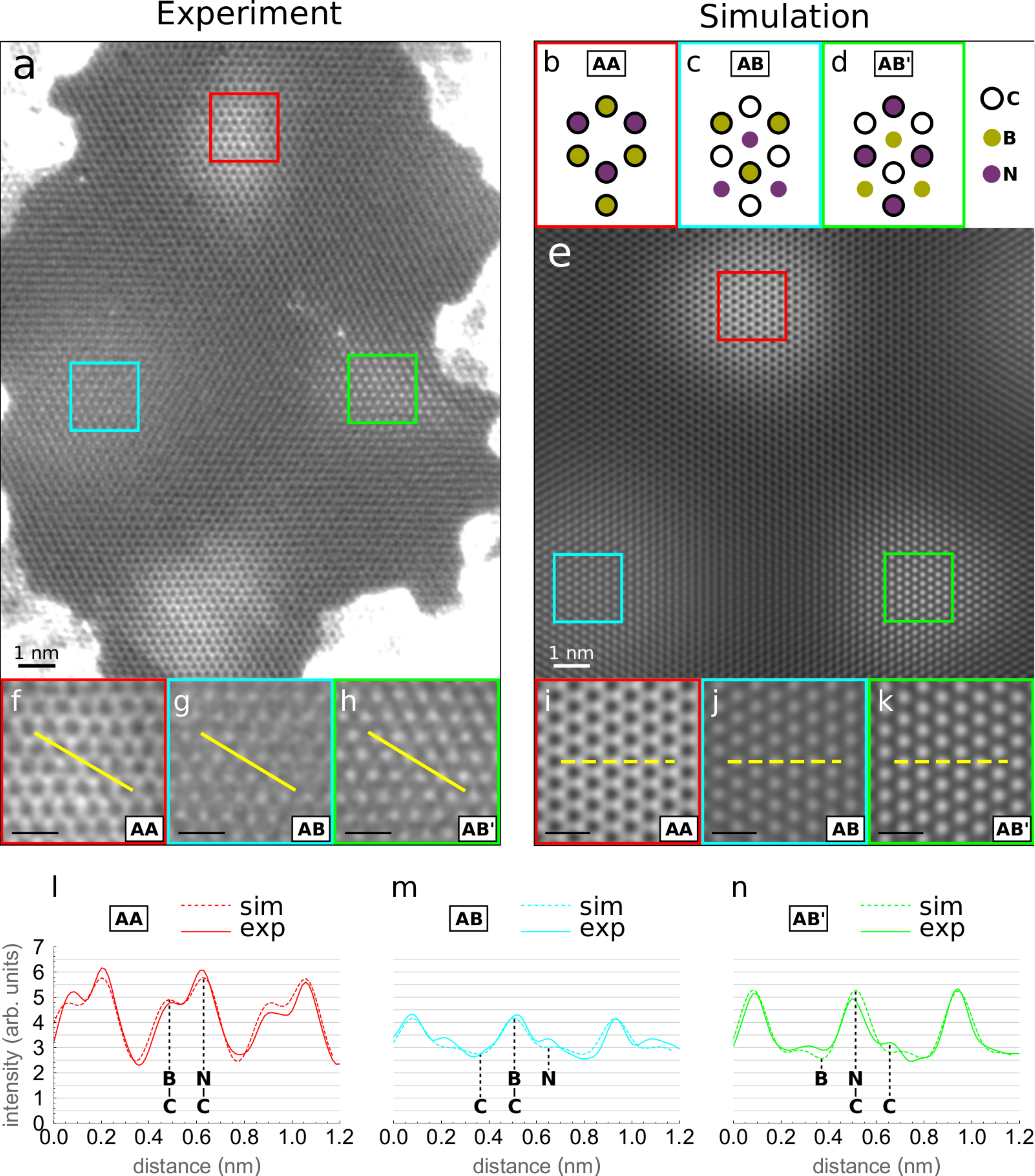}

\caption{MAADF imaging. (a) Atomically resolved MAADF image of a portion of
the heterostructure. Because of the contrast mechanism for medium
angle scattering, high-symmetry regions appear brighter. The top view
structure models of the high symmetry regions are shown in (b)-(d).
The regions in the colored squares of (a) are shown at higher magnification
in (f)-(h). (e) STEM MAADF simulation of the considered heterostructure.
(i)-(k) Magnified views of the three high-symmetry regions of (e).
(l)-(n) Gray value intensity profiles for the experimental (solid
lines) and the simulated (dashed lines) case along the yellow lines
in (f)-(k). Scale bars in panels (f)-(k) are all 0.5 nm.}

\label{maadf}
\end{figure}

\begin{figure}
\includegraphics[width=0.9\textwidth]{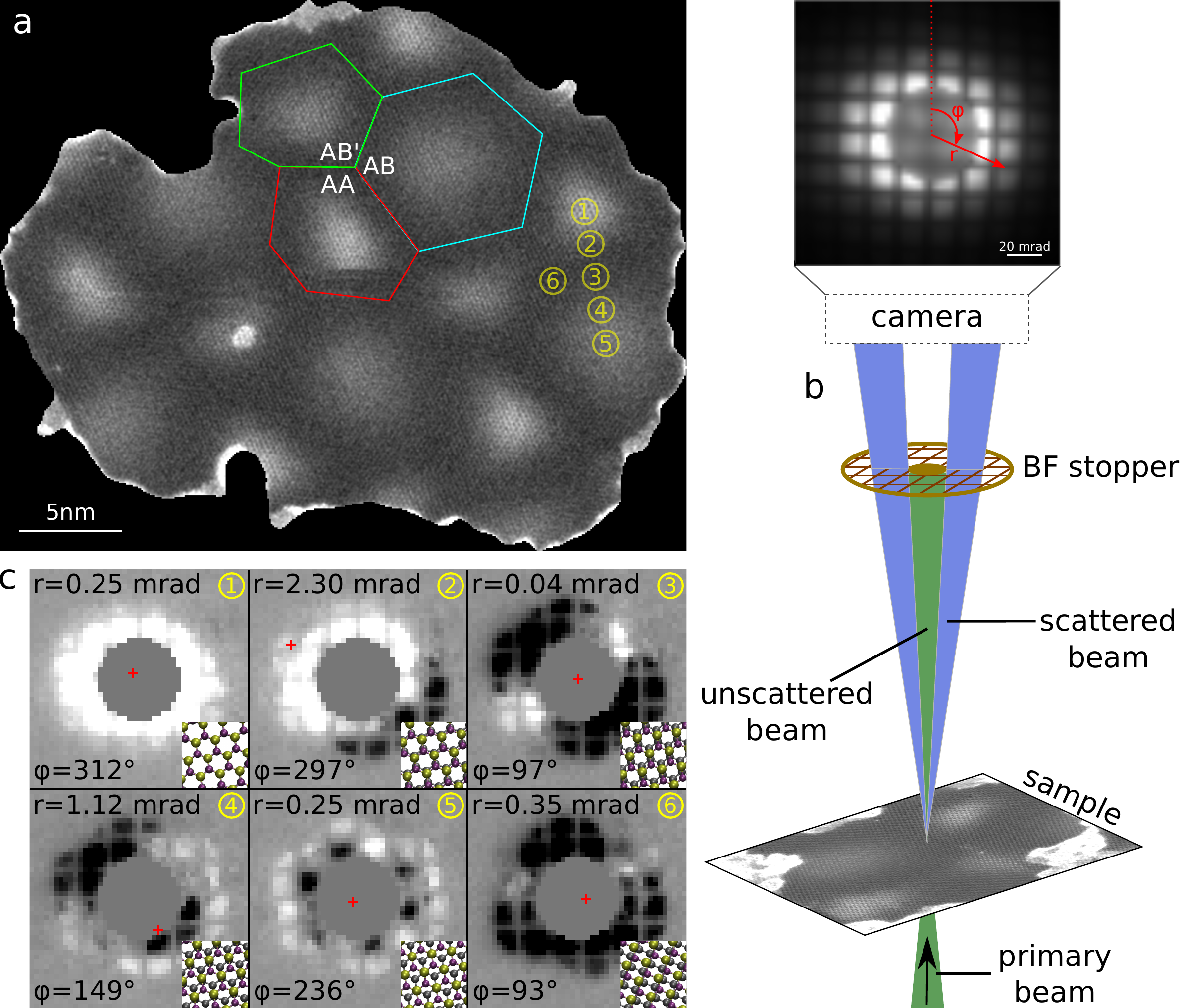}

\caption{(a) MAADF image of an area of the sample containing several moiré
regions. The AB stacked region (enclosed in the cyan polygon) is found
to be consistently larger than the AA (in red) and the AB' (in green)
regions. (b) Schematics of the experimental setup used for direction
sensitive detection of the scattered electrons, including an example
exposure recorded by the CCD camera and the polar coordinate system
used to describe the position of the ACOM. (c) From the indicated
regions (1-6) in (a), the scattering intensity distributions are shown
as the difference between a 10\texttimes 10 pixels area binned signal
and a reference signal that is obtained as an average of all recorded
images (excluding those corresponding to contamination). Insets illustrate
the local relative lattice offsets that are associated with the asymmetric
scattering intensity. The red cross indicates the position of the
ACOM in each image (the radial coordinate was exaggerated by a factor
of 20). The $r$ and $\varphi$ coordinates of the ACOM position are
also indicated for each image.}
\label{schematics}

\end{figure}

Further investigation on this sample was performed in a Nion UltraSTEM
100 operated at 60 kV. Fig.\ \ref{maadf}a shows an atomically resolved
medium angle annular dark field (MAADF) image of a small portion of
the suspended heterostructure. In the range of scattering angles used
here (ca. 60 to 200 mrad), regions where the atoms are precisely on
top of each other appear brighter because the intensity here is not
simply the sum of the two overlaid lattices, as it would be in a high
angle annular dark field (HAADF) image. \citep{Pennycook2011} Indeed,
in HAADF imaging (ca. 80 to 240 mrad) the intensity does not vary
across the differently stacked regions (see supplementary Fig.\ 1).
For this reason, and also because the MAADF image has a better signal-to-noise
ratio than the HAADF image, MAADF imaging was preferred over HAADF
in this work. There are three types of high-symmetry stacked regions
labeled as AA (C atoms aligned with B and N atoms), AB (C atoms aligned
with B atoms only) and AB' (C atoms aligned with N atoms only). The
top view structure models of the three stacking types are schematically
shown in Fig. \ref{maadf}b-d. In Fig.\ \ref{maadf}f-h the AA, AB
and AB' regions are shown at higher magnification in red, cyan and
green frames, respectively. A STEM simulation of the heterostructure,
performed using the QSTEM software,\citep{Koch2002} is shown in Fig.\ \ref{maadf}e
and the AA, AB and AB' regions are shown at higher magnification in
Fig.\ \ref{maadf}i-k. Fig.\ \ref{maadf}l-n show the intensity
profiles for each of the three regions along the yellow lines of Fig.\ \ref{maadf}f-k
for the experimental (solid line) and for the simulated (dashed line)
case. The AA region can be identified already from its visual appearance,
which is distinctly different from that of the AB and AB' regions
(compare Fig.\ \ref{maadf}f,i with Fig.\ \ref{maadf} g,h,j,k).
The AB and AB' region can be distinguished by comparing the intensity
modulation in the lattice, which is always stronger in the AB' region
(where C and N are aligned, Fig.\ \ref{maadf}n) than in the AB region
(where C and B are aligned, Fig.\ \ref{maadf}m). Hence, from the
appearance and relative intensity variations (Fig.\ \ref{maadf}l-n
are plotted with the same intensity scale), it is possible to unambiguously
associate each moiré spot to a specific stacking type. After careful
analysis of many regions across the sample, we observed that the AB
stacked regions consistently appear larger than the AA and AB' regions.
This can be clearly seen for instance in Fig.\ \ref{schematics}a,
where a MAADF image containing several moiré spots is presented (a
black mask was used here to cover contaminated areas). The three different
moiré regions were identified as explained above for Fig.\ \ref{maadf}.
Individual AA, AB and AB\textquoteright{} regions are enclosed by
red, cyan and green polygons respectively. The sides of the polygons
are placed approximately along the lines of minimum intensity between
two adjacent moiré regions. Already at a first glance, it is evident
that the AB region is the largest of the three. Indeed, as drawn in
Fig.\ \ref{schematics}a, the AB region measures 55 nm\textsuperscript{2},
while the AA region is 34 nm\textsuperscript{2}, and the AB' region
is 32 nm\textsuperscript{2}.

To get more insight into the local atomic stacking and to ultimately
understand the reason behind the different sizes of the three moiré
regions, we modified the detection mode of the STEM to extract information
on the preferential scattering direction of the electron beam after
interaction with the sample. A schematic drawing of the experimental
setup is shown in Fig.\ \ref{schematics}b. We used a charge-coupled
device (CCD) camera to record a two-dimensional image of the scattered
electron beam at every probe position during scanning, in a conceptually
similar way as reported in Ref.\ \citep{Pennycook2015,Muller-Caspary2016,Ophus2016},
with the substantial difference that the DF signal instead of the
BF disk was recorded in this experiment. The very intense bright field
disk was shielded using a custom made aperture consisting of a copper
disk placed on a conventional 3 mm TEM grid (see supplementary information
for more details on the custom aperture). An example of such an image
is shown in the upper part of Fig.\ \ref{schematics}b. The relation
between recorded image, preferential scattering direction and local
atomic stacking is based on the following argument: for high-symmetry
stacked regions, \textit{i.e.} at the center of AA, AB and AB' spots,
the heterostructure shows perfect in-plane isotropy and the electron
beam will be elastically scattered along a cone around the axis of
the primary beam. The corresponding image will therefore show symmetric
illumination with respect to its center. However, when the probe hits
the side of a moiré spot, where the two lattices are slightly off
register, the electrons experience an anisotropic potential that results
in the beam being predominantly scattered in one direction. Consequently,
the recorded image will show asymmetric illumination. Examples of
the locally obtained scattering intensity distributions are shown
in Fig.\ \ref{schematics}c, for selected points as drawn on the
MAADF image of Fig.\ \ref{schematics}a. The six images show important
differences: regions 1 and 5, respectively centered on the AA and
on the AB stacked regions, produce strong isotropic scattering of
the beam around the center of the detector, with the AA region being
the stronger scatterer of the two (as can also be seen from MAADF
images). Regions 2 and 4 are respectively selected slightly off the
centers of AA and AB spots and in the corresponding scattering images
the intensity is preferentially accumulated on one side of the detector.
Finally, the center of mass of the detected intensity for regions
3 and 6 is at the center of the image, but the signal shows a two-
and three-lobe geometry that mirrors the local symmetry of the corresponding
regions. Since this signal is very sensitive to the local (projected)
stacking of the two layers at each position, the comparison to simulated
data from model structures allows us to establish the 3D structure
of the free-standing bilayer heterostructure.

\begin{figure}
\includegraphics[width=0.8\textwidth]{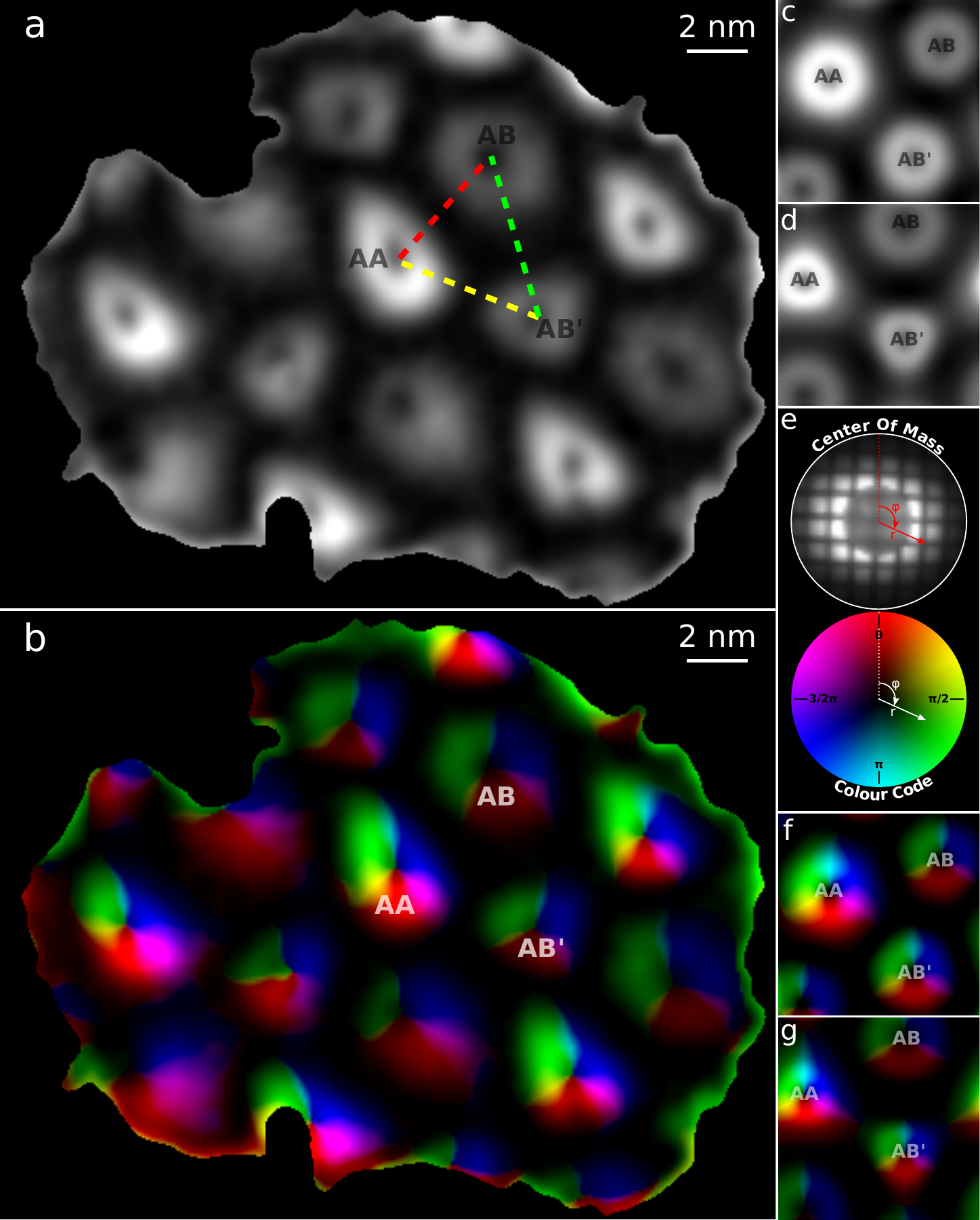}

\caption{Directional scattering analysis. (a) Radial ($r$) map of the ACOM
of the same area of Fig.\ \ref{schematics}a. The gray scale ranges
from black ($r=0$) to white ($r=r_{max}$). (b) Radial and angular
($r+\varphi$) map of the ACOM of the same area in (a). The color
of each pixel is assigned based on the position of the ACOM by a one
to one correspondence that is graphically explained in (e). (c) Simulated
$r$ map and (f) $r+\varphi$ map of the ACOM based on the rigid model.
(d) Simulated $r$ map and (g) $r+\varphi$ map of the ACOM based
on the relaxed model. Note that translations and rotations of the
maps must be allowed when comparing them to each other.}
\label{dir_scatt}
\end{figure}

\begin{figure}
\includegraphics[width=0.9\textwidth]{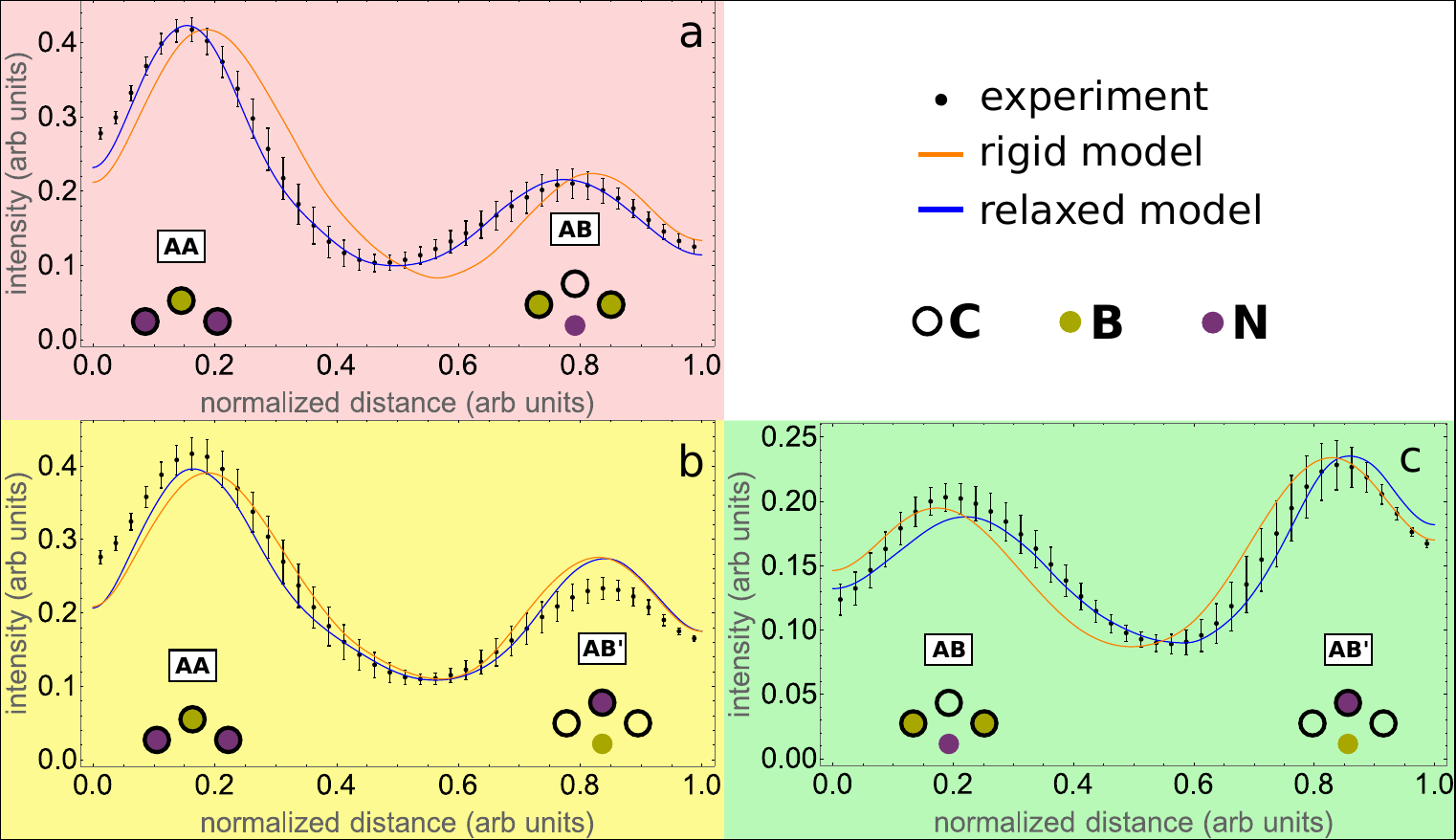}

\caption{Intensity profiles of the radial maps of Fig.\ \ref{dir_scatt} along
paths connecting AA to AB (a), AA to AB' (b) and AB to AB' (c). The
black dotted lines are the experimental profiles (obtained by averaging
7 to 12 individual profiles of Fig.\ \ref{dir_scatt}a), the orange
lines are the simulated profiles of the rigid model (Fig.\ \ref{dir_scatt}c)
and the blue lines are the simulated profiles of the relaxed model
(Fig.\ \ref{dir_scatt}d).}

\label{profiles}

\end{figure}

To quantitatively evaluate the preferential scattering direction we
take the diffracted intensity in the annular dark field pattern and
measure its center of mass (annular center of mass, abbreviated as
ACOM in the following) for each pixel of the scan. A representation
in polar coordinates ($r$ and $\varphi$) is used as schematically
illustrated in Fig.\ \ref{schematics}b. Examples of the calculated
$r$ and $\varphi$ values of the scattered intensities for the six
considered regions are noted in Fig.\ \ref{schematics}c. Here, the
position of the ACOM is also indicated by a red cross in each image,
showing that only for regions 2 and 5 the ACOM is significantly displaced
from the center of the detector. Fig.\ \ref{dir_scatt}a shows a
map of the same sample region of Fig.\ \ref{schematics}a obtained
by assigning to each pixel the $r$ value of the ACOM of the corresponding
diffracted intensity encoded by a gray scale, where black and white
colors correspond, respectively, to $r=0$ and $r=r_{max}$. As expected
from the considerations above, the map shows minima at the center
of the moiré spots, indicating perfect symmetry, and maxima around
these points, where the stacking offset produces prevalent electron
scattering in one direction. Note that points halfway between two
adjacent moiré spots are also dark. This is because in these regions
the atomic stacking is perfectly halfway between two high-symmetry
configurations and the coordinate $r$ of the ACOM goes to zero (see
regions 3 and 6 of Fig.\ \ref{schematics}c). Fig.\ \ref{dir_scatt}b
includes information on the angular direction of the preferential
scattering, where the coordinate $\varphi$ of the ACOM is encoded
by the color (see Fig.\ \ref{dir_scatt}e for graphical explanation
of the color code). It is interesting to observe how the scattering
direction depends on the angle around the center of a moiré spot,
spanning a range of 2$\pi$ around each. For comparison, we performed
STEM simulations based on a structure model consisting of a flat,
rigid graphene/hBN heterostructure. Saving the simulated exit waves
(ronchigrams) for each pixel allows us to treat the computed dataset
in the same way as its experimental counterpart. Fig.\ \ref{dir_scatt}c
and \ref{dir_scatt}f show the results of this simulation. The experimental
and the simulated maps show a qualitative agreement but important
differences become evident when comparing the relative sizes of the
moiré spots. The black dotted lines in Fig.\ \ref{profiles}a, b
and c are calculated by averaging 7 to 12 intensity profiles of Fig.\ \ref{dir_scatt}a
along straight paths connecting adjacent moiré spot centers. Three
such paths are indicated in Fig.\ \ref{dir_scatt}a by colored dashed
lines and they connect AA to AB (red), AA to AB' (yellow) and AB to
AB' (green). The position of the minimum in each of these plots, which
marks the transition between two adjacent stacking types, is significantly
different for the experiment and the simulation based on the rigid
model (orange solid lines). This disagreement can only be corrected
by considering a new structural model for the simulation that allows
for in-plane strain of the two crystals and/or out-of-plane distortion
of the heterostructure. The flat and rigid graphene/hBN model has
therefore to be abandoned in search of a more realistic atomic structure.

\begin{figure}
\includegraphics[width=1\textwidth]{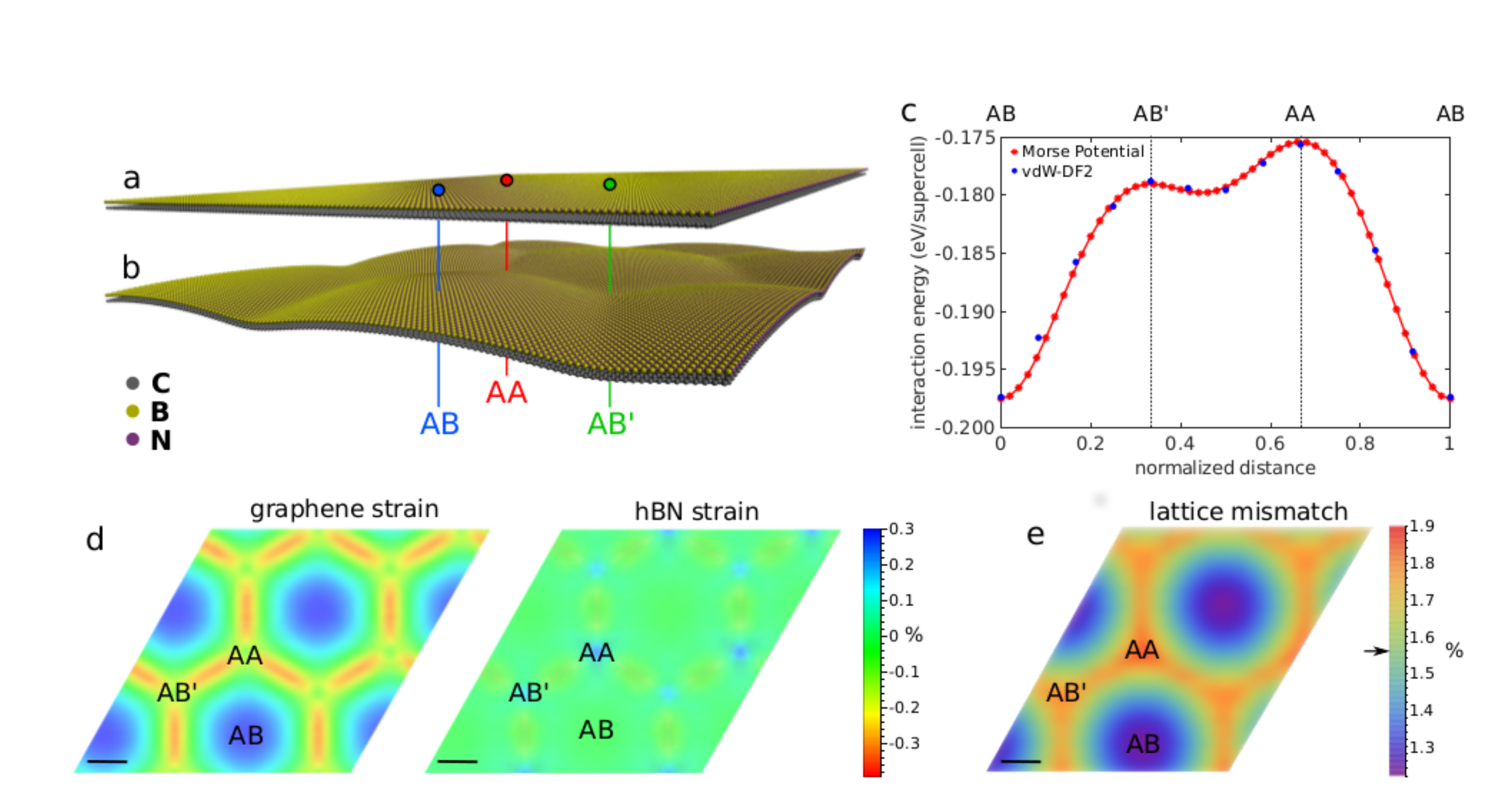}

\caption{Results of calculations. (a) Rigid structure model of the graphene/hBN
bilayer before relaxation. (b) Structure model of the graphene/hBN
bilayer after full relaxation. The relaxed model visibly distorts
in the out-of-plane direction. (c) Interlayer interaction energy plot
per supercell (four carbon, two boron and two nitrogen atoms). The
blue dots represent the values obtained by DFT calculations for different
stacking configurations, while the red dots indicate the shape of
the Morse potential, whose parameters were optimized to fit to the
DFT points. (d) In-plane strain maps of graphene (left) and hBN (right).
(e) Lattice mismatch map. The black arrow next to the color bar indicates
the initial lattice mismatch between the two crystals before the relaxation.
Scale bars in panels (d) and (e) are 2 nm.}
\label{calc}

\end{figure}

To this end we computed a relaxed graphene/hBN model by energy minimization
using a combination of density functional theory (DFT) calculations
and empirical potentials as explained in the following. Note that
the combination of the two methods is a key point here, since full-scale
DFT would be computationally prohibitive for a moiré unit cell consisting
of \ensuremath{\sim}16000 atoms, while empirical potentials have not
been reported so far for the case of graphene on hBN. In order to
determine the energy landscape of graphene on a hBN monolayer we followed
the same approach as in Ref.\ \citep{Zhou2015}, where several DFT
methods for simulating the van der Waals interaction between the two
layers were examined. For this work, the vdW-DF2 method \citep{Klimes2011,Thonhauser2007}
was preferred over the computationally expensive many-body adiabatic
fluctuation-dissipation theorem method \citep{Harl2010} and the DFT-D2
method, which accounts for long-range interactions through the addition
of a semi-empirical term.\citep{Grimme2006} A supercell consisting
of eight atoms (four carbon, two boron and two nitrogen atoms) was
constructed and we calculated the interaction energy between the two
layers defined as $E_{vdW}=E_{\infty}-E_{d_{0}}$, where $E_{\infty}$
and $E_{d_{0}}$ are the total energies of the supercell at infinite
and at the equilibrium interlayer distances, respectively. The blue
dots of Fig.\ \ref{calc}c show the calculated values of $E_{vdW}$
for the three high-symmetry stacking types and for other intermediate
disregistry configurations. The plot clearly shows that the AB type
is by far the most energetically favorable stacking type, followed
by AB' and finally by AA. This result is in agreement with existing
literature \citep{Sachs2011,Zhou2015} (note that in Ref.\ \citep{Zhou2015}
the AB and AB' structures were inadvertently misidentified, with their
names exchanged). To extend the calculation to the entire moiré unit
cell, we now describe the van der Waals interaction between the layers
by a Morse potential in the form $V(r)=D_{e}(e^{-2\alpha(r-r_{e})}-2e^{-\alpha(r-r_{e})})$,
where $D_{e}$ is the value of the potential at the equilibrium interlayer
distance $r_{e}$ and $\alpha$ sets the width of the potential. The
numerical values of the parameters were adjusted so that the interlayer
interaction agrees with the DFT results, leading to $D_{e}^{CB}=$
2.9 meV, $\alpha^{CB}=$ 2.08 Å\textsuperscript{-1} and $r_{e}^{CB}=$
3.86 Å for the C-B interaction and to $D_{e}^{CN}=$ 8.3 meV, $\alpha^{CN}=$
2.54 Å\textsuperscript{-1} and $r_{e}^{CN}=$ 3.84 Å for the C-N
interaction. With these values, an excellent match could be obtained
as shown by the red dots in Fig.\ \ref{calc}c. The C-C and B-N interaction
is treated using many-body lcbop \citep{Los2003} and Tersoff \citep{Tersoff1988,Tersoff1989}
potentials, respectively, leading to a lattice mismatch of \ensuremath{\sim}1.6
\%. Both potentials are implemented in the code \textit{large-scale
atomic/molecular massively parallel simulator} (LAMMPS).\citep{Plimpton1995,Plimpton2012}
For a moiré unit cell of graphene on hBN (0\textdegree{} misalignment),
65\texttimes 65 unit cells of graphene on 64\texttimes 64 unit cells
of hBN are needed to keep the periodic boundary conditions, totaling
16642 atoms. The structure with one degree of rotational misalignment
is avoided as one would need to consider millions of atoms to properly
model that structure. The total potential energy is minimized by relaxing
both layers without applying any constraint until the forces are below
$1.0\times10^{-6}$ eV/Å.

The initial and the fully relaxed models are presented in Fig.\ \ref{calc}a
and b, respectively, with the three stacking types marked. The relaxed
model visibly distorts in the out-of-plane direction, forming a wavy
structure with a periodicity that matches the moiré superlattice.
In particular, the AB region is found at a smooth bulge having the
concavity on the graphene side, while at AA and AB' regions the structure
has sharper kinks with the concavity facing the hBN side. The total
amplitude of the corrugation is \ensuremath{\sim}8.5 Å for each layer.
The results of our relaxation are in good agreement with the theoretical
prediction of Ref.\ \citep{Leven2016}.

STEM simulations based on the relaxed model were performed and the
resulting $r$ and $r+\varphi$ maps are shown in Fig.\ \ref{dir_scatt}d
and g, respectively. Although the rippling changes the positions of
the moiré spots, the distances and relative angles between them do
not change. Rather the main change to the moiré introduced by the
rippling is the size and shape of the spots. In particular the rippling
expands the AB region and causes the AA and AB\textquoteright{} regions
to become more triangular, in much better agreement with the experiment.
The intensity profiles for the map of the relaxed structure are shown
in Fig.\ \ref{profiles}a-c as blue solid lines. In Fig.\ \ref{profiles}a
and c the experimental and the simulated data based on the relaxed
model now show excellent agreement, with both central minima being
accurately reproduced. In other words, the plots from AA to AB (Fig.\ \ref{profiles}a)
and from AB to AB' (Fig.\ \ref{profiles}c) allow us to clearly distinguish
the rigid, flat model from the relaxed, rippled structure. Along the
line from AA to AB' (Fig.\ \ref{profiles}b), no significant difference
between the flat and rippled model can be identified. This is not
surprising because there is neither significant out of plane deformation
nor in-plane lattice distortion (discussed further below) along this
particular line in the relaxed structure. We also note that, while
the positions of the maxima and minima in the profiles for the experiment
and the simulation now match extremely well, the maximum amplitudes
of the $r$ values still deviate slightly (in Fig.\ \ref{profiles}b
and c the simulations slightly underestimate the amplitude near AA
and AB, while they are overestimated near AB'). This can not be due
to inaccuracies in the structure model, because distortions in the
membrane shape and layer alignment would shift the positions of maxima
and minima, but not affect their amplitude. Nonlinearities of the
detector, aberrations in the electron optics between sample and detector,
or remaining inaccuracies in modeling of the scattering might be the
reason. It is important to point out that none of these effects would
affect the locations of the minima in $r$ , since the minima reflect
special cases in the symmetry of the projected structure. At this
point, we will also comment briefly on how the ACOM analysis would
compare to analyzing the MAADF intensity: each of the ACOM profiles
as discussed above features two maxima and three minima, making the
position of the central minimum very sensitive to the transition point
between adjacent stacking types. A profile through the MAADF intensity,
on the other hand, only has two side maxima with a single broad central
minimum, which makes it very difficult to distinguish tiny differences
in the stacking transition.

The superior match between the experiment and the relaxed rippled
model indicates it is far more realistic than the flat model. We therefore
compute the strain maps for the two layers in the relaxed structure.
As shown in Fig.\ \ref{calc}d, the interatomic distances for both
graphene and hBN are not constant but modulated with a periodicity
matching the moiré wavelength. In particular, graphene tends to stretch
at AB regions and compress along lines connecting AA to AB' regions,
while hBN appears mostly unstrained, with small local stretching accumulated
at AA and AB' regions. From the strain distribution in each layer
we extract the lattice mismatch as shown in the map of Fig.\ \ref{calc}e.
Here it can be clearly seen how the two layers attempt to minimize
the mismatch at the AB regions, while at AA and AB' the mismatch is
significantly larger. This behavior can be explained by taking into
account two conflicting effects, as already discussed in Ref.\ \citep{Woods2014}.
As we demonstrated, AB is the most energetically favorable stacking
type and the two crystals will attempt, by a combination of stretching
and compression in each layer, to extend laterally this favorite stacking
and thus to gain in van der Waals energy, at the expense of the AA
and AB' regions that will necessarily shrink. This behavior is in
contrast to the elastic energy of the crystals' lattices, which scales
with the square of the strain and therefore attempts to restore the
intrinsic lattice constants. The equilibrium is reached when these
two competing forces cancel out. Note that the smallest value of lattice
mismatch is \ensuremath{\sim}1.2\%, indicating that the two lattices
are never found in a completely synchronous state.

In conclusion, we have presented here a TEM study of a free-standing
2D van der Waals heterostructure consisting of a well aligned bilayer
of graphene on hBN. Dark field imaging in a conventional TEM confirms
that the contaminants trapped between the two layers are squeezed
into a few isolated pockets, leaving most of the heterostructure with
an atomically clean interface. A direction sensitive acquisition mode
for the scattered electron beam of a STEM was developed and employed
to extract in-depth information on the local atomic stacking. Comparison
with STEM simulations based on a relaxed model indicates that the
heterostructure corrugates in the out-of-plane direction, with an
undulation having the same periodicity as the moiré pattern and a
total amplitude (in each layer) of \ensuremath{\sim}8.5 Å. This work
shows that depending on lattice mismatch and stacking misorientation,
suspended heterostructures, usually regarded as pure 2D materials,
should be effectively considered as 3D objects, with van der Waals
interlayer forces playing a key role in determining the in-plane strain
and out-of-plane deformation of each layer.
\begin{acknowledgement}
G.A., A.M., C.M., C.K., and J.C.M.\ acknowledge funding from the
European Research Council (ERC) Project No.\ 336453-PICOMAT.\ M.R.A.M.\ and
J.C.M.\ acknowledge financial support from the FWF Project No.\ P25721-N20.
J.K.\ acknowledges funding from the Wiener Wissenschafts- Forschungs-
und Technologiefonds (WWTF) via project MA14-009. T.J.P.\ acknowledges
funding from the European Union\textquoteright s Horizon 2020 research
and innovation programme under the Marie Sk\l odowska-Curie grant
agreement No.\ 655760 - DIGIPHASE. Computational resources from the
Vienna Scientific Cluster are gratefully acknowledged. A.K.G. acknowledges
funding from the European Research Council (ERC) project ARTIMATTER.
\end{acknowledgement}

\vspace{2cm}
\noindent
\textbf{Supporting Information}

\noindent
A comparison between MAADF and HAADF imaging, additional information on the custom aperture, as well as details on the STEM simulations are available in the Supporting Information.
\vspace{2cm}
\bibliographystyle{unsrt}
\bibliography{bibliography}

\end{document}